\documentclass[twocolumn,showpacs,preprintnumbers,eqsecnum]{revtex4}

\usepackage{amsmath}
\usepackage{makeidx}
\usepackage{graphicx}
\usepackage{dcolumn}
\usepackage{bm}
\usepackage{hyperref}

\begin{document}

\preprint{APS/123-QED}
\title{The total cross-sections for the photoeffect for K-shell bound
electrons and pair production with the created electron in the ground state\\
for photon energies above 1 MeV }
\author{A. Costescu$^{1}$,S. Spanulescu$^{1,2}$,C. Stoica$^{1}$}
\affiliation{$^{1}$Department of Physics, University of Bucharest, MG11,
Bucharest-Magurele 76900,Romania\\
$^{2}$Department of Physics, Hyperion University of Bucharest, Postal code
030629, Bucharest, Romania}

\begin{abstract}
Considering the contributions of the main term of the relativistic
Coulombian Green function given by Hostler to the second order of S matrix
element and taking into account only the large components of the Dirac
spinor of the ground state we obtain the imaginary part of the Rayleigh
amplitudes in terms of elementary functions. Thereby simple and high
accurate formulae for the total cross-sections for photoeffect and pair
production with the electron created in the K-shell are obtained \textit{via}
the optical theorem.Comparing the predictions given by our formulae with the
full relativistic numerical calculations of Kissel \textit{et al} [Phys.
Rev. A 22, 1970 (1980)] and Scofield [LLRL, Internal Report,1973], a good
agreement is found for photon energies above the pair production threshold
up to 5 MeV for any Z elements. We present our numerical results for the
total photoeffect and pair production cross-sections, for various photon
energies for the K-shell of Ag and Pb.
\end{abstract}

\maketitle

\section{\label{sec:level1}Introduction}

The S matrix element for Rayleigh scattering of an initial photon with
momentum $\overrightarrow{k_{1}}=\omega \overrightarrow{\nu }_{1}$and
polarization vector $\overrightarrow{s}_{1}$, and a final photon with
momentum $\overrightarrow{k_{2}}=\omega \overrightarrow{\nu }_{2}$and
polarization vector $\overrightarrow{s}_{2}$, by a K-shell bound electron,
if we take into account only the large components of the Dirac spinor of the
ground state and only the main term of the relativistic coulombian Green
function\cite{ho}, is:\cite{cs}\vspace{-0.3 cm}

\begin{equation}
\mathcal{M}_{R}=M\left( \omega ,\theta \right) (\overrightarrow{s}_{1}%
\overrightarrow{s}_{2})+N\left( \omega ,\theta \right) (\overrightarrow{s}%
_{1}\overrightarrow{\nu }_{2})(\overrightarrow{s}_{2}\overrightarrow{\nu }%
_{1})  \label{M}
\end{equation}

where $\theta $ is the photon scattering angle, and the invariant amplitudes 
$M$ and $N$ are:

\begin{equation}
M(\omega ,\theta )=\mathcal{O}-P\left( \Omega _{1},\theta \right) -P\left(
\Omega _{2},\theta \right)  \label{am}
\end{equation}

\begin{equation}
N(\omega ,\theta )=Q\left( \Omega _{1},\theta \right) +Q\left( \Omega
_{2},\theta \right)  \label{an}
\end{equation}

with the Rayleigh scattering amplitudes given by:

\begin{equation}
P\left( \Omega ,\theta \right) =\frac{2^{7}\lambda ^{5}X^{3}}{d^{4}(\Omega )}%
\frac{\omega \pm \omega _{pp}}{2m}\frac{F_{1}(2-\tau ;2,2;3-\tau
;x_{1},x_{2})}{2-\tau }  \label{ap}
\end{equation}

\begin{equation}
Q\left( \Omega ,\theta \right) =\frac{2^{11}\lambda ^{5}X^{3}\omega ^{2}}{%
d^{5}(\Omega )}\frac{\omega \pm \omega _{pp}}{2m}\frac{F_{1}(3-\tau
;3,3;4-\tau ;x_{1},x_{2})}{3-\tau }  \label{aq}
\end{equation}

where\vspace{-0.5 cm}

\begin{eqnarray*}
\Omega _{1}\hspace{-0.2cm} &=&\hspace{-0.2cm}\gamma m+\omega ,\Omega
_{2}=\gamma m-\omega =-\left\vert \Omega _{2}\right\vert , \\
\gamma &=&\left( 1-\alpha ^{2}Z^{2}\right) ^{\frac{1}{2}},\lambda =\alpha
Zm,\omega _{pp}=(1+\gamma )m, \\
\tau _{j}\hspace{-0.2cm} &=&\hspace{-0.2cm}\frac{\alpha Z\Omega _{j}}{X_{j}}%
,X_{j}=m^{2}-\Omega _{j}^{2}\text{ with Re[}X_{j}]>0,j=1,2
\end{eqnarray*}

The function $\mathcal{O}$ is the atomic form factor, while $F_{1}\left(
a;b_{1},b_{2};c;x_{1},x_{2}\right) $ is the Appell hypergeometric function
of four parameters and two complex variables given by the relationships: 
\vspace{-0.3cm} 
\begin{equation}
x_{1}x_{2}=p=\left[ \frac{d^{\ast }\left( \Omega _{1}\right) }{d\left(
\Omega _{1}\right) }\right] ^{2}=\xi ^{2}  \label{sum}
\end{equation}
\vspace{-0.6cm} 
\begin{equation}
x_{1}+x_{2}=s=2\xi -\frac{16X^{2}\omega ^{2}\sin ^{2}\frac{\theta }{2}}{%
d^{2}\left( \Omega _{1}\right) }=\xi ^{2}  \label{prod}
\end{equation}
\vspace{-0.8cm} with \vspace{-0.3cm} 
\begin{eqnarray}
d\left( \Omega _{j}\right) &=&2m(\pm \gamma \omega -\alpha ^{2}Z^{2}m+\alpha
ZX_{j}),  \label{d} \\
d^{\ast }\left( \Omega _{j}\right) &=&2m(\pm \gamma \omega -\alpha
^{2}Z^{2}m-\alpha ZX_{j})  \notag
\end{eqnarray}

In the equations (\ref{ap})-(\ref{aq}) and (\ref{d}) the upper sign
corresponds to the case $\Omega =\Omega _{1}$, while the lower sign
corresponds to the case $\Omega =\Omega _{2}.$

Obviously, in the case of the forward scattering, the amplitude $N(\omega
,\theta )$ is no longer present in the Rayleigh amplitude. Also, the
imaginary part of the amplitude $\mathcal{M}_{R}$ is given only by the terms 
$P\left( \Omega _{1},0\right) $ and $P\left( \Omega _{2},0\right) $ which
are present in the expressions of the photoeffect and the pair production
cross-sections respectively.

\vspace{-0.3cm}

\section{The total cross-section of the photoelectric effect and pair
production in the case of K shell}

According to the optical theorem, the imaginary part of the Rayleigh
amplitude for forward scattering allows to get the total photoeffect
cross-section \textit{per} K-shell electron:

\vspace{-0.3 cm}

\begin{widetext}%

\begin{equation}
\sigma _{ph}=\frac{4\pi }{\alpha }\frac{m}{\omega }r_{0}^{2}\left\vert \text{%
Im[}P\left( \Omega _{1},0\right) ]\right\vert =\frac{16\pi ^{2}}{3}%
r_{0}^{2}m^{2}\alpha ^{5}Z^{6}\frac{\omega +\omega _{pp}}{2m}\frac{\Omega
_{1}\left\vert X_{1}^{2}\right\vert }{\omega ^{5}}\left( 1+\left\vert \tau
_{1}\right\vert ^{2}\right) \frac{\left( -\xi _{1}\right) ^{\tau _{1}}}{%
e^{\pi \left\vert \tau _{1}\right\vert }-e^{-\pi \left\vert \tau
_{1}\right\vert }}  \label{sph}
\end{equation}

In a similar way we get the total pair production cross-section \textit{per} K-shell electron:

\begin{equation}
\sigma _{pp}=\frac{4\pi }{\alpha }\frac{m}{\omega }r_{0}^{2}\left\vert \text{%
Im[}P\left( \Omega _{2},0\right) ]\right\vert =\frac{16\pi ^{2}}{3}%
r_{0}^{2}m^{2}\alpha ^{5}Z^{6}\frac{\omega -\omega _{pp}}{2m}\frac{%
\left\vert \Omega _{2}\right\vert \left\vert X_{2}^{2}\right\vert }{\omega
^{5}}\left( 1+\left\vert \tau _{2}\right\vert ^{2}\right) \frac{\left( -\xi
_{2}\right) ^{\tau _{2}}}{e^{\pi \left\vert \tau _{2}\right\vert }-e^{-\pi
\left\vert \tau _{2}\right\vert }}  \label{spp}
\end{equation}
\end{widetext}

We point out that: $\left( -\xi _{1}\right) ^{\tau _{1}}=e^{-\left\vert \tau
_{1}\right\vert \chi _{1}}$, with 
\begin{equation}
\chi _{1}=\pi -2\arctan \left( \frac{\alpha Z\left\vert X_{1}\right\vert }{%
\gamma \omega -\alpha ^{2}Z^{2}m}\right) ;\omega >\frac{\alpha ^{2}Z^{2}m}{%
\gamma }  \label{hi1}
\end{equation}

and $\left( -\xi _{2}\right) ^{\tau _{2}}=e^{-\left\vert \tau
_{2}\right\vert \chi _{2}}$, with 
\begin{equation}
\chi _{2}=\pi -2\arctan \left( \frac{\alpha Z\left\vert X_{2}\right\vert }{%
\gamma \omega +\alpha ^{2}Z^{2}m}\right) ;\omega >\omega _{pp}  \label{hi2}
\end{equation}

\section{Numerical results and conclusions}

\begin{figure*}[tbp]
\includegraphics[width=4.in]{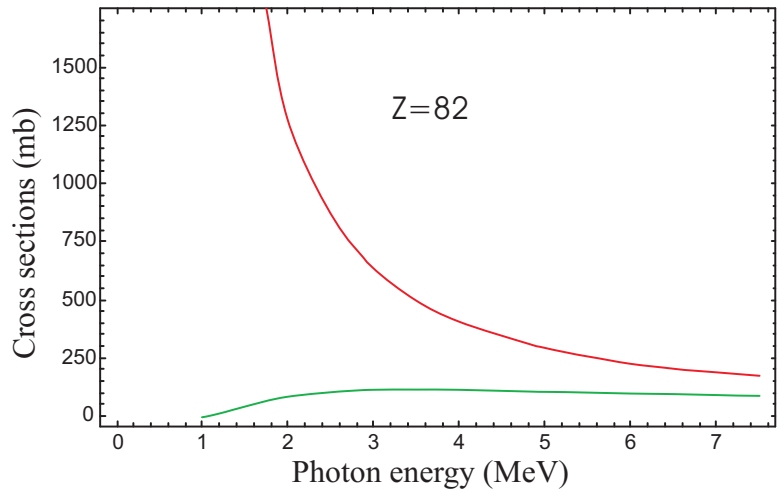}
\caption{Photoeffect (red) and pair production (green) cross sections for Pb
K-shell}
\label{figura1}
\end{figure*}

Using our analytical formulae for the cross sections for K-shell electrons
we get the numerical numerical results in Table \ref{Ag}, Table \ref{Pb},
and figure \ref{figura1} for the whole K-shell.

\begin{widetext}%

\begin{table*}[h]%

\caption
{Photoeffect and pair production cross sections for Ag K-shell.}%
\label{Ag}%
\begin{tabular}{r l  l  r @{.}   l}
\hline\hline Energy (keV) \hspace*{10mm}
& Pair production &     Photoeffect &
\multicolumn{2}{l} {Cross sections}\\
\hspace{10mm} & cross section (mb) \hspace*{10mm}& cross section (mb)\hspace
*{10mm}&\multicolumn{2}{l}{ ratio}\\
\hline1000 \hspace*{10mm}  &  \texttt{~}9.145x$10^{-7}$    &  493.836 & 5&399x$10^{8}$\\
1100  \hspace*{10mm}  & \texttt{~}0.0875335 &  401.376   & 4585&39\\
1249  \hspace*{10mm}  &  \texttt{~}0.851678&   306.868  &  360&309  \\
1500  \hspace*{10mm}  &  \texttt{~}3.11888&    211.705 &    67&878\\
2000  \hspace*{10mm}  &   \texttt{~}7.16782 &  122.626   &   17&107   \\
2754  \hspace*{10mm}  &  \texttt{~}9.96235 &  \texttt{~}70.2819   &   7&054  \\
3000  \hspace*{10mm}  &   10.3218 &   \texttt{~}61.0716 &   5&916   \\
3500  \hspace*{10mm}  &  10.6112  &  \texttt{~}47.8107  & 4&505  \\
4000  \hspace*{10mm}  &   10.5516 &  \texttt{~}38.99   &   3&695    \\
4807  \hspace*{10mm}  &  10.1208  &  \texttt{~}29.7828  &  2&942    \\
5000  \hspace*{10mm}  &  \texttt{~}9.98901 &  \texttt{~}28.1578  &  2&818     \\
5500  \hspace*{10mm}  &  \texttt{~}9.62702 &  \texttt{~}24.6344   & 2&558     \\
6000  \hspace*{10mm}  & \texttt{~}9.25496  &  \texttt{~}21.8602   &  2&362   \\
6500  \hspace*{10mm}  &  \texttt{~}8.88765  &  \texttt{~}19.6248   &  2&208     \\
7000  \hspace*{10mm}  &  \texttt{~}8.53296 &  \texttt{~}17.7887   &  2&084    \\
7500  \hspace*{10mm}  &  \texttt{~}8.1948  &   \texttt{~}16.2559  &  1&983    \\
\hline\end{tabular}%
\end{table*}%


\begin{table*}\centering%
\caption
{Photoeffect and pair production cross sections for Pb K-shell.}%
\label{Pb}%
\begin{tabular}{r l  l  r @{.}   l}
\hline\hline Energy (keV) \hspace*{10mm}
& Pair production &     Photoeffect &
\multicolumn{2}{l} {Cross sections}\\
\hspace{10mm} & cross section (mb) \hspace*{10mm}& cross section (mb)\hspace
*{10mm}&\multicolumn{2}{l}{ ratio}\\
\hline 1000 \hspace*{10mm}  &  \texttt{~~}0.11370   &  5099.07      & 44846&5\\
1100  \hspace*{10mm}  &  \texttt{~~}2.36329 &     4151.89 &      1756&82\\
1249  \hspace*{10mm}  &  \texttt{~}12.2336 &     3181.11 &      260&03\\
1500  \hspace*{10mm}  &  \texttt{~}37.1725 &  2200.44   &  59&195   \\
2000  \hspace*{10mm}  &  \texttt{~}79.4739 &  1278.86 &    16&091  \\
2754  \hspace*{10mm}  & 107.831 &   \texttt{~}735.099&   6&817   \\
3000  \hspace*{10mm}  & 111.338&     \texttt{~}639.19  &  5&741  \\
3500  \hspace*{10mm}  & 113.924&     \texttt{~}500.947&    4&397  \\
4000  \hspace*{10mm}  & 112.949 &    \texttt{~}408.876   & 3&619   \\
4807  \hspace*{10mm}  & 108.012 &   \texttt{~}312.65   & 2&894    \\
5000  \hspace*{10mm}  & 106.549 &   \texttt{~}295.652  &  2&774 \\
5500  \hspace*{10mm}  & 102.571 &   \texttt{~}258.781  &  2&522  \\
6000  \hspace*{10mm}  &   \texttt{~}98.517&   \texttt{~}229.731  &  2&331 \\
6500  \hspace*{10mm}  &    \texttt{~}94.5392 &    \texttt{~}206.311  &  2&182\\
7000  \hspace*{10mm}  &   \texttt{~}90.7119  &  \texttt{~}187.065     & 2&062 \\
7500  \hspace*{10mm}  &   \texttt{~}87.0731 &  \texttt{~}170.993  & 1&963 \\
\hline\end{tabular}%
\end{table*}%
\end{widetext}

\vspace{2.3 cm}As it may be observed from equations (\ref{hi1}) and (\ref%
{hi2}), for higher photon energies the arguments of the exponentials, $\chi
_{1}$ and $\chi _{2}$, have the same limit, $\pi-2\arctan[\frac{\alpha Z }{%
\gamma}]$, so that the two cross sections will become closer and closer as
the photon energy increases.

Comparing the predictions given by our formulae with the full relativistic
numerical calculations of Kissel \textit{et al} \cite{ki} and Scofield\cite%
{Sco}, a good agreement is found for photon energies above the pair
production threshold up to 5 MeV for any Z elements.

The good agreement of our calculation with the full relativistic results
shows that, for the presented energies regime, the main relativistic
kinematics terms are canceled by some retardation and multipols terms, and
the spin effects are small.

\vspace*{0.5cm}

\section*{ACKNOWLEDGEMENT}

This work was partially supported by the Romanian National Research
Authority (ANCS) under Grant CEEX PC-D11-PT00-582/2005.

\end{document}